\documentclass[aps,pra,twocolumn,showpacs,floatfix,superscriptaddress]{revtex4}
\usepackage{graphicx}
\begin{document}

\title{Prospects for sympathetic cooling of molecules \\
in electrostatic, ac and microwave traps}

\author{S. K. Tokunaga}
\affiliation{Centre for Cold Matter, Blackett Laboratory,
Imperial College London, Prince Consort Road, London SW7 2AZ, United Kingdom.}
\author{Wojciech Skomorowski}
\affiliation{Faculty of Chemistry, University of Warsaw,
Pasteura 1, 02-093 Warsaw, Poland}
\author{Piotr S. {\.Z}uchowski}
\affiliation{Department of Chemistry, Durham University,
South Road, Durham DH1 3LE, United Kingdom}
\author{Robert Moszynski}
\affiliation{Faculty of Chemistry, University of Warsaw,
Pasteura 1, 02-093 Warsaw, Poland}
\author{Jeremy M. Hutson}
\affiliation{Department of Chemistry, Durham University,
South Road, Durham DH1 3LE, United Kingdom}
\author{E. A. Hinds}
\author{M. R. Tarbutt}
\affiliation{Centre for Cold Matter, Blackett Laboratory,
Imperial College London, Prince Consort Road, London SW7 2AZ, United Kingdom.}
\begin{abstract}
We consider how trapped molecules can be sympathetically cooled
by ultracold atoms. As a prototypical system, we study LiH
molecules co-trapped with ultracold Li atoms. We calculate the
elastic and inelastic collision cross sections of $^7$LiH +
$^7$Li with the molecules initially in the ground state and in
the first rotationally excited state. We then use these cross
sections to simulate sympathetic cooling in a static electric
trap, an ac electric trap, and a microwave trap. In the static
trap we find that inelastic losses are too great for cooling to
be feasible for this system. The ac and microwave traps confine
ground-state molecules, and so inelastic losses are suppressed.
However, collisions in the ac trap can take molecules from
stable trajectories to unstable ones and so sympathetic cooling
is accompanied by trap loss. In the microwave trap there are no
such losses and sympathetic cooling should be possible.
\end{abstract}

\pacs{37.10.Mn, 34.50.-s, 34.50.Cx}

\maketitle

\section{Introduction}\label{intro}

There has been rapid progress in the field of cold and
ultracold molecular gases over the last decade, driven by a
diverse range of applications in physics and chemistry
\cite{Carr09}. Polar molecules are of particular interest
because they interact strongly with applied electric fields,
and interact with one another through dipole-dipole
interactions that are long-range, anisotropic, and tuneable.
These properties, along with the exceptional control that is
possible at low temperatures over all the degrees of freedom,
make an ultracold gas of polar molecules an ideal tool for
simulating strongly interacting condensed-matter systems and
the remarkable quantum phenomena they exhibit \cite{Micheli06}.
In low-temperature molecular gases it becomes possible to
control chemical reactions using electric and magnetic fields
and to study the role of quantum effects in determining
chemical reactivity \cite{Krems:2005}. Cold molecules are also
useful for testing fundamental symmetries, for example by
measuring the value of the electron's electric dipole moment
\cite{Hudson:2002}, searching for a time-variation of
fundamental constants \cite{Hudson:2006, Bethlem:2008,
Bethlem:2009}, or measuring parity violation in nuclei
\cite{DeMille:2008} or in chiral molecules \cite{Darquie:2010}.
For these applications, a great leap in sensitivity could be
obtained by cooling the relevant molecules to low temperatures
so that, for example, the experiment could be done in a trap or
a fountain \cite{Tarbutt(2):2009}.

The bialkali molecules can be produced at very low temperatures
by binding together ultracold alkali atoms, by either
photoassociation \cite{Sage05, Deiglmayr08} or
magnetoassociation \cite{Ospelkaus08, Lang08, Danzl08,
Danzl10}. A few specific species of other molecules are
amenable to direct laser cooling to ultralow temperatures
\cite{Shuman10}. A large variety of useful molecules can be
produced with temperatures in the range 10\,mK to 1\,K by
decelerating supersonic beams \cite{Bethlem99, Narevicius08,
Fulton06} or capturing the lowest-energy molecules formed in a
cold buffer-gas source \cite{Weinstein98, vanBuuren09}. For
many applications it is desirable to cool these molecules to
lower temperatures, and this could be done by mixing the
molecules with ultracold atoms and encouraging the two to
thermalize. This sympathetic cooling method has not yet been
demonstrated for neutral molecules, but is often used to cool
neutral atoms \cite{Myatt97, Truscott01, Modugno01}, atomic
ions \cite{Larson86} and molecular ions \cite{Drewsen04,
Ostendorf97}.

For sympathetic cooling to yield ultracold molecules, the rate
of atom-molecule elastic collisions, which are responsible for
the cooling, must be sufficiently high that the molecules cool
in the available time. In practice this requires that both
atoms and molecules be trapped, so that they are held at high
density and interact for a long time. The easiest way to trap
molecules is in a static electric or magnetic trap. However,
static traps can confine molecules only in weak-field-seeking
states; since the ground state is always strong-field-seeking,
inelastic collisions can eject molecules from the trap by
de-exciting them to lower-lying strong-field-seeking states.
These traps are therefore unsuitable for sympathetic cooling
unless the ratio of the elastic to inelastic cross section
happens to be particularly large. Inelastic losses can be
avoided by trapping ground-state molecules, but such traps are more
difficult to realize.

In this paper, we consider the sympathetic cooling of LiH
molecules with ultracold Li atoms. Due to its large dipole
moment of 5.88\,D, its low mass, and its simple structure, LiH
is an attractive molecule for studying the physics of dipolar
gases and the electric field control of collisions and chemical
reactions. A supersonic beam of cold LiH molecules has been
produced \cite{Tokunaga07} and decelerated to low speed using a
Stark decelerator \cite{Tokunaga09}. Ultracold Li is likely to
be a good coolant for LiH because the closely matching masses
ensures that energy is transferred efficiently in an elastic
collision. Also, the low mass of Li ensures that inelastic
collisions with non-zero angular momentum are suppressed by a
centrifugal barrier, even at relatively high collision energies
\cite{Lara06}. We have prepared a magneto-optical trap of
$10^{10}$ Li atoms for the purpose of sympathetic cooling.

We begin by calculating the elastic and inelastic cross
sections for LiH + Li collisions. Then we calculate the
trajectories of a set of trapped LiH molecules that have
occasional collisions with a co-trapped cloud of ultracold Li.
Our aims are to calculate how the molecular temperature evolves
with time, to investigate loss mechanisms in different kinds of
traps, and to establish how the ultracold atoms should be
distributed so that the cooling is most efficient.

\section{Scattering Calculations}

We have carried out quantum-mechanical scattering calculations
on $^7$Li+$^7$LiH collisions on the potential energy surface of
ref.\ \onlinecite{Skomorowski:LiHLi-pot:2010}. The calculations
are carried out using the MOLSCAT program \cite{molscat:v14}.
We use full close-coupling calculations for the energy range of
importance for sympathetic cooling, up to collision energies of
1~K, and coupled states (CS) calculations over an extended
range up to 100~cm$^{-1}$. The calculations are carried out
treating LiH as a rigid rotor, with rotational constant $b_{\rm
LiH}/hc=7.5202$ cm$^{-1}$. Because of the deep potential well
(8743 cm$^{-1}$) and strong anisotropy, a large rotational
basis set is needed. The present calculations include all
functions with LiH rotational quantum number $j$ up to $j_{\rm
max}=37$. The coupled equations are solved using the hybrid
log-derivative/Airy propagator of Alexander and Manolopoulos
\cite{Alexander:1987} with the propagation continued to 500\
\AA.

The collision calculations treat the Li atom as structureless.
This is justified because there are almost no terms in the
collision Hamiltonian that can cause a change in the Li
hyperfine state or magnetic projection quantum number
\cite{Soldan:MgNH:2009, Zuchowski:NH3:2009}. The (very small)
hyperfine structure of the LiH molecule is also neglected.

\begin{figure}[tbh]
\begin{center}
\includegraphics[width=0.33\textwidth,angle=-90]{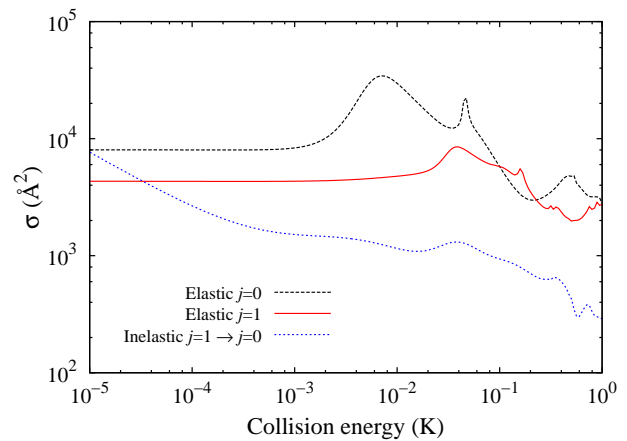}
\caption{(color online)
Elastic (red) and inelastic (blue) cross sections from
close-coupling calculations on $^7$Li-$^7$LiH with $^7$LiH
initially in $j=1$. Also shown (black) are elastic cross sections for
$^7$LiH initially in $j=0$.} \label{fig:cc}
\end{center}
\end{figure}

The results of close-coupling calculations for LiH molecules
initially in $j=1$ are shown in Figure \ref{fig:cc}. As
expected from the Wigner threshold laws \cite{Wigner:1948}, the
elastic cross section becomes constant at very low energy
(below about 1~mK) and the inelastic cross section is
approximately proportional to $E^{-1}$ in this region. Above
about 10~mK the ratio of elastic to inelastic cross sections
stabilizes at a factor of 5 to 10.

\begin{figure}[tbh]
\begin{center}
\includegraphics[width=0.33\textwidth,angle=-90]{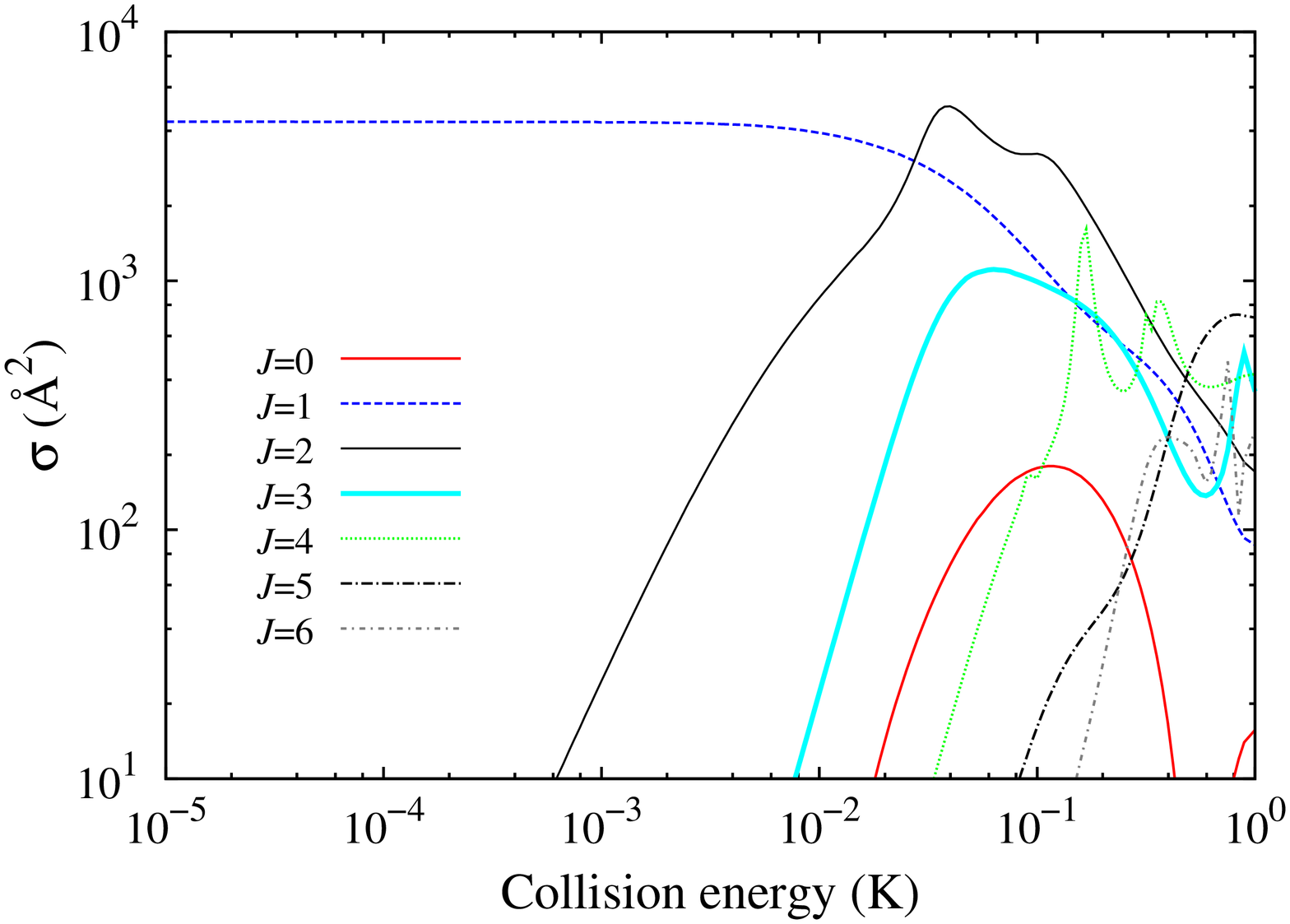}
\includegraphics[width=0.33\textwidth,angle=-90]{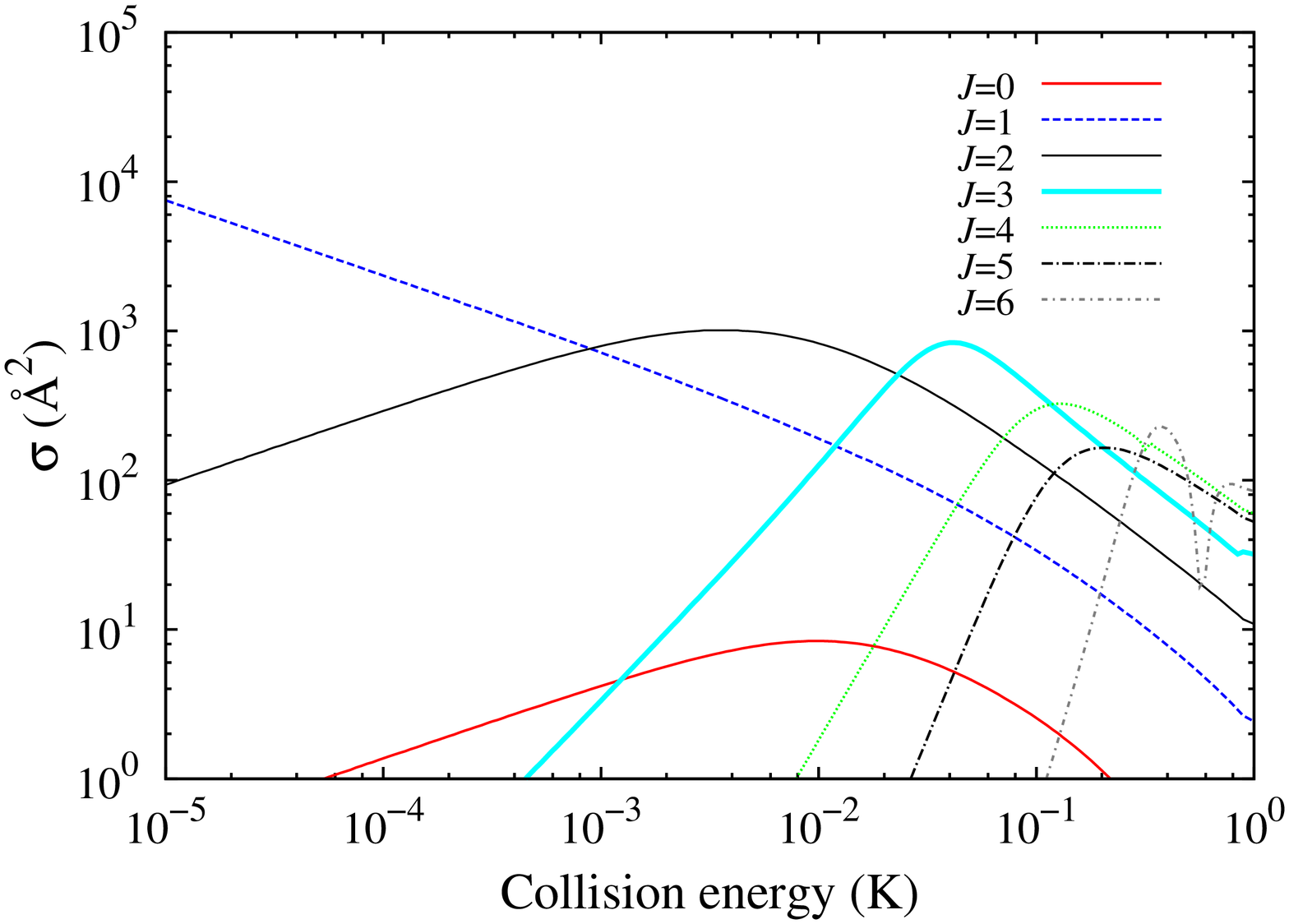}
\caption{(color online)
Partial-wave contributions to elastic (upper panel) and
inelastic (lower panel) cross sections from close-coupling
calculations.} \label{fig:cc-part}
\end{center}
\end{figure}

Close-coupling calculations are carried out for fixed values of
the total angular momentum $J$, and the resulting partial wave
contributions are summed to form cross sections. The results in
Figure \ref{fig:cc} include contributions up to $J_{\rm
max}=10$, and Figure \ref{fig:cc-part} shows the individual
partial wave contributions for $0<J\le6$. There is no
significant resonance structure in the inelastic cross sections
for $J<6$, although shape resonances appear in the elastic
cross sections for $J\ge2$ and are particularly prominent for
$J=4$ and 6.

The cross sections obtained from quantum scattering
calculations on an individual potential energy surface are in
general quite sensitive to small potential scalings because of
variations in the scattering length $a$. However, this
sensitivity is much smaller in Li+LiH because of the relatively
low reduced mass. The low-energy limit of the elastic cross
sections shown in Figure \ref{fig:cc} may be compared with the
value $\overline\sigma=4\pi\overline a^2=3200$~\AA$^2$ obtained
from the mean scattering length $\overline a=16.01$~\AA, as
defined by Gribakin and Flambaum \cite{Gribakin:1993}.

\begin{figure}[htb]
\begin{center}
\includegraphics[width=0.33\textwidth,angle=-90]{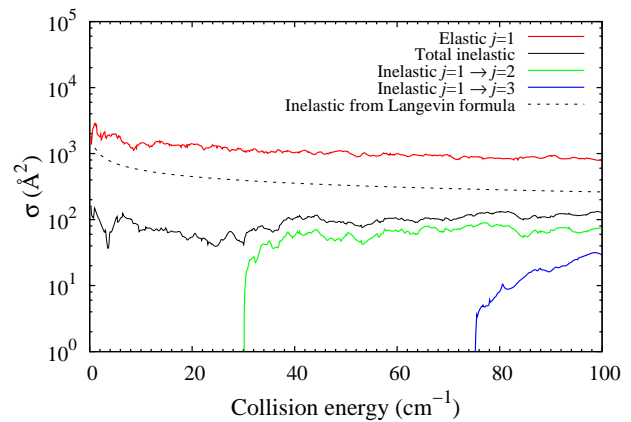}
\caption{(color online)
Elastic (red) and state-to-state inelastic cross sections from
coupled states calculations on $^7$Li-$^7$LiH with $^7$LiH
initially in $j=1$. The dashed black line shows the Langevin
limit for the total inelastic cross section.} \label{fig:cs}
\end{center}
\end{figure}

The higher-energy results from CS calculations are shown in
Fig.\,\ref{fig:cs}. It is interesting to compare the inelastic
cross section with the Langevin limit, which assumes that all
collisions that cross the centrifugal barrier lead to inelastic
events. The Langevin limit is shown as a dashed line in
Fig.\,\ref{fig:cs}, and it may be seen that the inelastic cross section
remains below this limit even at collision energies around
100~cm$^{-1}$.

We note that the reaction $\text{LiH + Li} \rightarrow \text{Li}_{2} + \text{H}$ is highly endothermic and so cannot occur at the low collision energies of interest here.

\section{Sympathetic cooling simulations}

Using the cross sections calculated in Section II, we simulate
the sympathetic cooling of LiH molecules co-trapped with
ultracold Li atoms. We consider three types of trap. The first
is a static electric trap for molecules in the
weak-field-seeking state $(j,m)=(1,0)$. Here, elastic
collisions with the Li atoms cool the molecules, whereas an
inelastic collision transfers a molecule to a lower-lying
high-field-seeking state causing it to be lost from the trap.
The other two traps we consider are a microwave trap and an ac
electric trap, both of which can trap ground-state molecules so
that inelastic losses are avoided. We note that sympathetic cooling in an optical dipole trap has been studied previously \cite{Barletta2010}.

\subsection{Cooling in a static electric trap}

We first consider the sympathetic cooling of LiH molecules in
the weak-field-seeking $(j,m)=(1,0)$ state. The simulation
starts with a large set of molecules with a velocity
distribution that fills the trap. Later, when we consider ac
and microwave traps, we will track individual molecular
trajectories in these traps, but we do not need do this for the
electrostatic trap. Our aim is only to calculate the fraction
of all the molecules that cool to low temperature without being
lost from the trap and this is determined entirely by the ratio
of elastic to inelastic cross sections as a function of the
collision energy.

For each collision we find the kinetic energy in the centre of mass frame, look up the
relative probability for elastic and inelastic collisions as
given in Figure \ref{fig:cc}, and then make a random choice
between these two processes according to this probability. If
the collision is inelastic, the molecule is lost from the trap.
If the collision is elastic, the molecule's velocity is
transformed by the collision into a new velocity using a
hard-sphere collision model. The velocity vector of the atom is
selected at random from an isotropic Gaussian velocity
distribution whose width is fixed by the temperature of the
atom cloud, chosen here to be 140\,$\mu$K. The atom and
molecule velocities are transformed into the centre of momentum
frame, where the molecular momenta before and after the collision,
$\mathbf{p}$ and $\mathbf{p'}$, are related by

\begin{equation}\label{reflection}
\mathbf{p'} = \mathbf{p} - 2\left( \mathbf{p} \cdot
\hat{\mathbf{e}} \right)\hat{\mathbf{e}},
\end{equation}

\noindent where $\hat{\mathbf{e}}$ is a unit vector along the
line joining the centres of the spheres. It is given by

\begin{equation}\label{lineCentres}
\hat{\mathbf{e}} = \sqrt{1-|\mathbf{b}|^{2}}
\frac{\mathbf{p}}{|\mathbf{p}|} + \mathbf{b}
\end{equation}

\noindent where $\mathbf{b}$ is a vector that lies in the plane
perpendicular to $\mathbf{p}$ and whose magnitude is the impact
parameter of the collision normalized to the sum of the radii
of the two spheres. For each collision, $\mathbf{b}$ is chosen
at random from a uniform distribution subject to the
constraints $\mathbf{b} \cdot \mathbf{p} = 0$ and $|\mathbf{b}|
\le 1$. The new momentum of the molecule, $\mathbf{p'}$, is
finally transformed back into the lab frame and this momentum
is used in the next collision.

\begin{figure}[t]
\includegraphics[width=8cm]{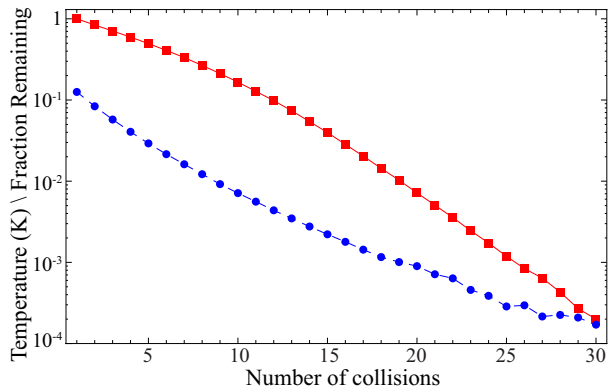}
\caption{(Color online)
Cooling and loss in a trap for weak-field-seeking molecules.
The temperature of the molecules (blue circles) and the
fraction remaining in the trap (red squares) are plotted
against the number of collisions.} \label{TempAndFrac}
\end{figure}

Figure \ref{TempAndFrac} gives the results of these simulations, showing how the fraction of molecules
remaining in the trap and the temperature of their distribution
depends on the number of collisions that have occurred. The
molecules have an initial temperature near 100\,mK, and for
every 10 collisions their temperature falls by about a factor
of 10. After about 20 collisions, they have reached a
temperature of 1\,mK, but only 0.7\% of them remain. After 28
collisions, the molecules have thermalized to the temperature
of the atoms, but now the fraction that remains is only
$4\times 10^{-4}$. We see that the ratio of elastic to
inelastic cross sections is too small in this case for
sympathetic cooling in a static trap to be feasible.

Our simulations use collision cross sections calculated in zero field, even though the molecules are electrostatically trapped. An electric field can have a large effect on atom-molecule collisions, as recently demonstrated for collisions between Rb and ND$_{3}$ in an electrostatic trap \cite{Parazzoli11}. Here, it was found that the trapping field increases the inelastic cross section. If a similar effect occurred for Li-LiH, it would strengthen our conclusion that sympathetic cooling is not feasible in the static trap for this system.

\subsection{Cooling in an ac electric trap} \label{sec:actrap}

To eliminate trap loss due to inelastic collisions, it is
desirable to trap the molecules in their ground state. The
ground state of every molecule is strong-field-seeking, and
strong-field-seeking molecules cannot be trapped using static
fields. One solution is to use an ac electric trap where the
molecules move on a saddle-shaped potential that focusses them
towards the centre of the trap in one direction, but defocusses
them in another direction. By alternating the focussing and
defocussing directions at a suitable rate, molecules are
confined near the trap centre. Such ac electric traps have
already been used to trap polar molecules in
strong-field-seeking states \cite{vanVeldhoven05, Bethlem06,
Schnell07}, and also to trap ground-state atoms
\cite{Schlunk07, Rieger07}, and sympathetic cooling of
molecules in ac traps has been proposed \cite{Schlunk07,
Rieger07}.

\begin{figure}[t]
\includegraphics[width=8.5cm]{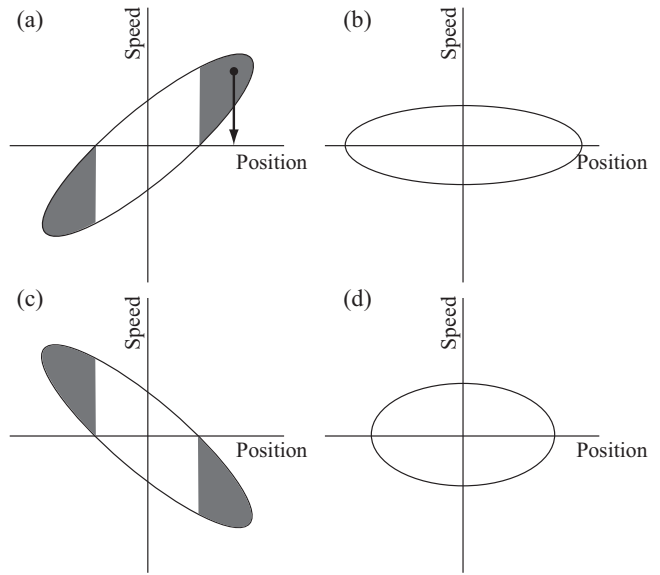}
\caption{
Phase-space acceptance of an ac trap in one dimension for 4
phases of the switching cycle: (a) start of focussing period,
(b) centre of focussing period, (c) start of defocussing
period, (d) centre of defocussing period. The shaded areas
indicate regions of phase space that are unstable in an
idealised head-on collision.} \label{Ellipses}
\end{figure}

Motion in an ac trap consists of a small-amplitude micromotion
at the switching frequency of the trap, superimposed on a
larger-amplitude, lower-frequency macromotion. The stability of
the molecules derives from the micromotion, and a collision
which interrupts the micromotion may put the molecule onto an
unstable trajectory. This means that a molecule initially
confined in the trap may be ejected by a collision even though
the collision reduces its energy. Figure \ref{Ellipses} gives a
simple picture of how this can happen. In each dimension, the
set of all the stable molecules forms an ellipse in phase
space, and this ellipse evolves periodically with the phase of
the switching cycle \cite{Bethlem06}. Figure \ref{Ellipses}(a)
shows the ellipse at the start of a focussing phase, showing
that the positions and speeds of the molecules are positively
correlated at this phase. To illustrate what may happen in a
collision, consider the case where molecules collide head-on
($\mathbf{b}=0$) with stationary atoms of the same mass. In
this idealised case, a collision reduces the speed of a
molecule to zero, leaving its position unchanged, as indicated
by the arrow in Figure \ref{Ellipses}(a). The molecule will
remain trapped only if it is still inside the ellipse, so the
shaded regions of the ellipse are unstable against collisions.
The same arguments apply at the start of the defocussing phase,
as indicated in Figure \ref{Ellipses}(c). Half-way through the
focussing and defocussing phases (Figure \ref{Ellipses}(b,d)),
the molecules remain inside the stable region when their speeds
are reduced to zero and so there will be no collisional loss at
these phases. The phase-space plots show what happens in only
one dimension. In a cylindrically symmetric ac trap, the same
plots can be made for the radial and longitudinal directions
separately, one being half a period out of phase with the
other. Thus when the focussing phase begins in the radial
direction, the defocussing phase is beginning in the
longitudinal direction, and a molecule must not be inside any
of the shaded regions if it is to remain trapped after a
collision.

This picture is, of course, a highly simplified one. The
collisions do not reduce the speed to zero, and they can couple
energy from one direction to another, which tends to increase
the opportunities for loss. Nevertheless, we expect the same
general conclusions to hold -- a large portion of the trap's
stable phase-space volume becomes unstable when sympathetic
cooling collisions are introduced, and collisions are more
likely to result in loss at the start of a focus/defocus phase
than half-way through.

Turning now to a complete simulation, we consider LiH molecules
in a cylindrical ac trap consisting of two ring electrodes and
two cylindrically symmetric end caps, as used in refs.\
\cite{vanVeldhoven05, Bethlem06}. The square of the electric
field magnitude in this trap is well approximated by the
expression

\begin{eqnarray}\label{eFieldAC}
E^{2}(z,\rho) = E_{0}^{2}
\left(1 + 2a_{3}\frac{(z^{2}-\frac{1}{2}\rho^{2})}{z_{0}^{2}} +
a_{3}^{2}\frac{(z^{4}+\frac{1}{4}\rho^{4})}{z_{0}^{4}}
\right.\nonumber \\* \left. +
2a_{5}\frac{(z^{4}-3z^{2}\rho^{2}+\frac{3}{8}\rho^{4})}{z_{0}^{4}}\right),
\end{eqnarray}

\noindent where $z_{0}$ is the characteristic size of the trap,
$E_{0}$ is the electric field magnitude at the trap centre, and
$a_{3}$ and $a_{5}$ are the coefficients in a multipole
expansion of the electrostatic potential. Considering the same
trap as used in ref.\ \cite{Bethlem06}, we set $a_{3}=-1.29$
and $a_{5}=0.63$ for the longitudinal focussing phase,
$a_{3}=1.29$ and $a_{5}=0.44$ for the radial focussing phase,
$z_{0}=4.55$\,mm, and $E_{0}=50$\,kV/cm. We switch the trap
between the two configurations at a frequency of 5\,kHz with a
50:50 duty cycle.

An ensemble of initially warm molecules evolves within the
trap, each molecule having occasional collisions with a
distribution of $10^{10}$ ultracold Li atoms. We calculate the
trajectories of many molecules moving in the trap by solving
the equations of motion numerically using a Runge-Kutta method
with a fixed time step. The force acting on the molecules is
$\mathbf{F}=-\mathbf{\nabla} W$ where $W$ is the Stark shift.
For the electric field magnitudes considered here, the Stark
shift of ground-state LiH is small compared to the rotational
spacing, and is given to a good approximation by second-order
perturbation theory:

\begin{equation}\label{WJM}
W=- \frac{\mu_{e}^{2} E^{2}}{6 b_{\rm LiH}}.
\end{equation}
Here, $\mu_{e}$ is the electric dipole moment of the molecule
and $b_{\rm LiH}$ is the rotational constant (in energy units).

We suppose that the atoms are trapped independently from the
molecules, for example in a magnetic trap, and we give them a
spherically symmetric Gaussian spatial distribution with a
$1/e$ half-width $w_a = 3$\,mm, and a Maxwell-Boltzmann
velocity distribution with a temperature $T_a=50\,\mu$K.

We first simulate trajectories without any collisions, starting
with an initial phase-space distribution that is larger than
the trap acceptance, so as to obtain a set of molecules that,
in the absence of collisions, survive in the trap for 10\,s.
This set of molecules defines the phase-space acceptance of the
trap and is then used for the full simulation including the
collisions. This ensures that molecules are lost from the trap
only as a result of collisions. After each interval of time
$\Delta t$, the calculation of the molecular trajectory is
stopped and the probability, $P$, of the molecule having a
collision during this time interval is calculated. The value of
$\Delta t$ is chosen such that $P \ll 1$ and we take $P = n \,
\sigma(K) \, v \, \Delta t$, where $n$ is the local density of
atoms, $v$ is the relative velocity of the LiH molecule and Li
atom at the time of collision, and $\sigma(K)$ is the elastic
collision cross section at collision energy $K$. A random
number, $r$, is chosen from a uniform distribution in the
interval from 0 to 1, and a collision occurs only if $P>r$.
When a collision does occur, the velocity vector obtained from
the molecular trajectory is transformed using the hard-sphere
collision model outlined above. The numerical integration of
the trajectory then continues using this transformed velocity.
Since the number of trapped atoms is many orders of magnitude
larger than the number of trapped molecules, we assume that the
atom distribution is unaffected by the presence of the
molecules. We also neglect collisions between the molecules,
since their density is so low.

\begin{figure}[t]
\includegraphics[width=8.5cm]{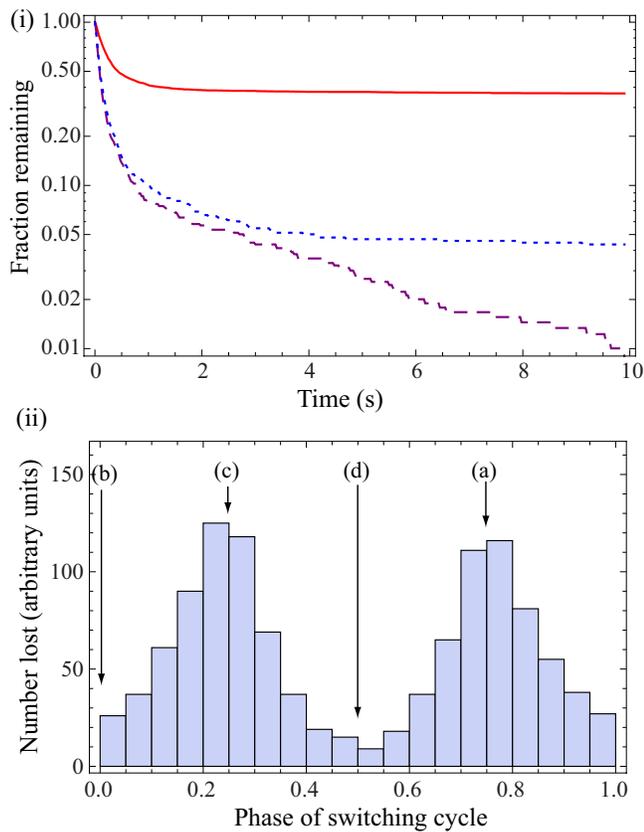}
\caption{(Color online)
Simulated molecule loss in an ac trap. (i) Fraction of molecules
surviving as a function of time.  Dashed (purple) line: real
trap, $T_a=50\,\mu$K. Dotted (blue) line: real trap, $T_a=0$.
Solid (red) line: ideal trap, $T_a=50\,\mu$K. (ii) Number of
collisions that result in molecule loss as a function of the
phase of the switching cycle, in an ideal trap. The zero of
phase corresponds to the centre of the radial focussing period
and the labels (a-d) refer to the phases depicted by the same
labels in Figure \ref{Ellipses}.} \label{acLoss}
\end{figure}

The dashed line in Figure \ref{acLoss}(i) shows the fraction of
molecules that survive in the ac trap as a function of time. We
see that most of the molecules are lost due to collisions and
that this loss occurs on two separate time scales. During the
first 1\,s, 94\% of all the molecules are lost from the trap.
Between 1\,s and 10\,s the number of trapped molecules
continues to fall, so that after 10\,s only 1\% remain in the
trap. It is surprising to find that the loss continues at long
times since we would expect there to be a small region close to
the origin of phase space that is stable against collisions. It
appears that even these molecules are eventually being
destabilized by the collisions. To investigate why this
happens, we repeated the simulation with the atom temperature
reduced to zero. The result is shown by the dotted line in Fig.
\ref{acLoss}(i). Here, the loss at early times is the same as
before, but after a few seconds of cooling the fraction
remaining in the trap stabilizes at around 5\%. When the atoms
have non-zero temperature, the collisions cause molecules near
the origin of phase-space to diffuse away from the origin,
eventually ending up on an unstable trajectory. The atoms cool
the hotter molecules, but they also tend to heat the coolest
ones, and even when the atom temperature is only $50\,\mu$K the
heating results in significant additional losses from the trap
on a 10\,s timescale.

Next, to shed some light on why there is so much collisional
loss in the ac trap, we simplified the simulations by
neglecting terms in the electric field beyond the second term
in Eq.\,(\ref{eFieldAC}). In this harmonic approximation the
phase-space acceptance of the trap is maximized, and although
this field cannot be realized in practice \cite{Bethlem06} it
is helpful to make this approximation since the dynamics in
such an ideal ac trap are well understood. The higher-order
terms complicate the dynamics by introducing nonlinear forces
into the trap and coupling the axial and radial motions, and
this greatly reduces the trap acceptance. Our simulations show
that neglect of these higher-order terms increases the
acceptance by a factor of 4 in both position and velocity. The
solid line in Figure \ref{acLoss}(i) shows the fraction of
molecules that survive in the ideal ac trap as a function of
time. In this case almost all the loss occurs in the first 1\,s
of cooling. The loss occurs from the outer regions of the trap
and there is a `safe' region around the phase-space origin
where a molecule will remain inside the trap's acceptance for
all possible outcomes of a single collision. Once molecules
have been cooled into this region there are no losses. The
fraction of all the initial molecules that remain in the trap
after 3\,s is 38\%, and there are no further losses between
3\,s and 10\,s. These results conform to the intuitive
expectations obtained from our discussion of Figure
\ref{Ellipses}. Comparing the results obtained for the ideal
trap with those of the real trap, we see that it is primarily
the large non-linear forces that make the ac trap an unsuitable
environment for sympathetic cooling.

Figure \ref{acLoss}(ii) shows how the number of collisions
resulting in trap loss depends on the phase of the switching
cycle, in the ideal ac trap. As expected from the discussion of
Figure \ref{Ellipses}, the simulations confirm that collisions
occurring at the start of a focussing / defocussing period are
far more likely to cause loss than those occurring half-way
through these periods. Similar results are obtained when the
higher-order terms are included, except that the modulation
observed in the figure is then not so deep.

\begin{figure}[t]
\includegraphics[width=8.5cm]{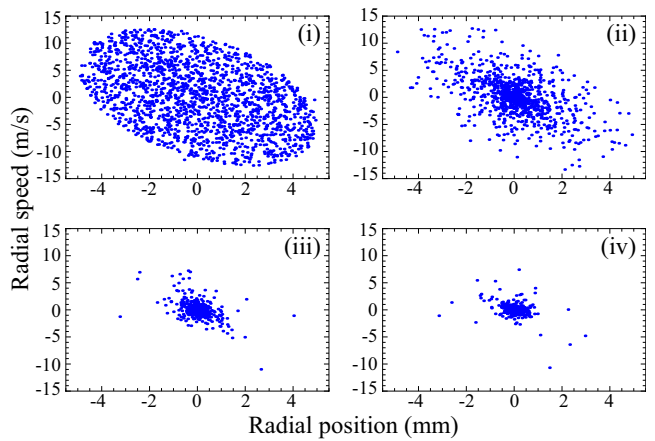}
\caption{(Color online)
Simulated time evolution of the radial phase-space distributions of
molecules in an ideal ac trap overlapped with a 50\,$\mu$K atom
cloud with a width parameter of $w_a=3$\,mm. Only the 1st and
2nd terms in Eq.\,(\ref{eFieldAC}) are included in the
expression for the electric field. The phase of the switching
cycle is the same in each plot and, to within one switching
cycle, the times are (i) 0.002\,s, (ii) 0.5\,s, (iii) 3\,s and
(iv) 10\,s.} \label{acPSDs}
\end{figure}

Figure \ref{acPSDs} shows how the radial phase-space
distribution in the ideal ac trap evolves with time. At early
times the molecules fill the available trap acceptance, but as
time goes on they congregate near the origin of phase space.
After 0.5\,s [Figure \ref{acPSDs}(ii)] the ellipse has become
dense near the centre and sparse elsewhere. Most molecules have
had one or more collisions by this time and these collisions
tend to remove molecules from the outer regions of the
distribution, either by cooling them towards the centre or, as
discussed above, kicking them out of the trap. As time goes on,
almost all the molecules in the outer regions of phase space
disappear and the molecules that remain are cooled into a small
region near the origin. After 10\,s [Figure \ref{acPSDs}(iv)]
this cold distribution has a full width at half maximum of
0.3\,mm in radial position and 0.9\,m/s in radial speed. The
time evolution of the longitudinal phase-space distribution is
similar.

\subsection{Cooling in a microwave trap} \label{sec:mctrap}

An alternative way to trap ground-state molecules is to use a
microwave trap, as discussed in ref.\ \cite{DeMille04}. The
ground-state molecules are attracted to the electric field
maximum of the standing-wave microwave field inside a resonant
cavity. The trap depth is particularly large when the detuning
of the microwave frequency from the rotational transition
frequency is small, although this places a stringent
requirement that the microwave field be circularly polarised in
order to avoid multi-photon excitation to rotationally excited
states \cite{DeMille04}. Collision-induced absorption of
microwave photons may also occur in the trap, and again this
unwanted process is far more probable when the detuning is
small \cite{Alyabyshev09}. Here, we consider a far-detuned
microwave trap for ground-state LiH molecules, operating at a
frequency of 15\,GHz. Since this frequency is very small
compared to the rotational frequency ($2b_{\rm
LiH}/h=445$\,GHz), and since the Stark shift will also be small
compared to $b_{\rm LiH}$ for all attainable electric field
strengths, the Stark shift is given to a good approximation by
Eq.\,(\ref{WJM}), where $E^{2}$ is now the time-averaged
squared electric field. We take the microwave field to be the
fundamental Gaussian mode of a symmetrical Fabry-Perot cavity,
having a beam waist of 15\,mm and an rms electric field at the
trap centre of $E_{0} = 40$\,kV/cm. This is the field produced
by coupling 2.6\,kW of power into a cavity whose Q-factor is
set by the reflectivity of room-temperature copper mirrors. The
trap has a depth of 500\,mK and the simulation begins with a
trap whose phase-space acceptance is completely filled. We
simulate individual molecular trajectories in the microwave
trap with collisions modelled in exactly the same way as
outlined above for the ac trap. We use the field-free elastic cross section shown in Fig.\,\ref{fig:cc} since this is insensitive to the microwave field \cite{Alyabyshev09}, and we neglect inelastic relaxation between field-dressed states because, for our trap conditions, the rate of this inelastic process is expected to be very low compared to the elastic collision rate \cite{Alyabyshev09}.

Each time a molecule has a collision with an ultracold atom,
its energy is reduced. Nevertheless, it is possible for a
collision to transfer energy between axial and radial motions
so that it has enough energy in one direction to leave the
trap. By running simulations both with
and without collisions, we find that there is no
additional trap loss as a result of the collisions. This is because the axial and radial motions in the trap are
weakly coupled, so that even in the absence of collisions a molecule whose energy is greater than the trap depth eventually leaves the trap.

\begin{figure}[t]
\includegraphics[width=8.5cm]{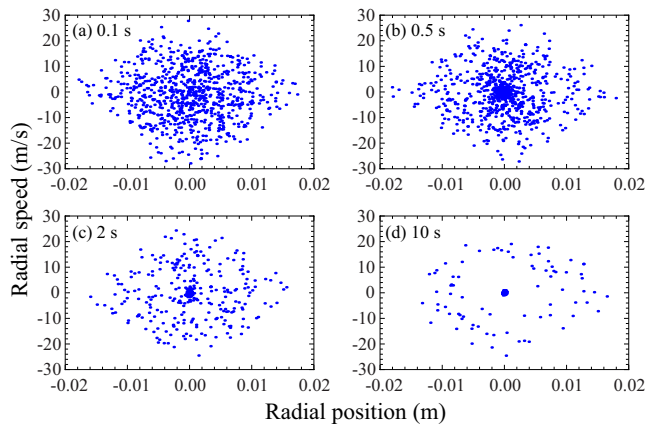}
\caption{(Color online)
Simulated time evolution of the radial phase-space distributions of
molecules in the microwave trap overlapped with a 140\,$\mu$K
atom cloud with a width of $w_a=3$\,mm. The cooling times are
(a) 0.1\,s, (b) 0.5\,s (c) 2\,s and (d) 10\,s.}
\label{microwavePSDs}
\end{figure}

Figure \ref{microwavePSDs} shows how the distribution of
molecules in the trap evolves with time as they cool to the
temperature of the atom cloud whose width is $w_a=3$\,mm. Each
plot shows the radial position and speed of each molecule in
the trap. After 0.1\,s very little cooling has occurred and the
molecules have the full range of speeds and positions that the
trap can accept. After 0.5\,s it is clear that the molecules
are accumulating near the phase-space origin as expected. They
have small speeds and are confined near the centre of the trap.
As time goes on, the accumulation of cold molecules continues.
After 2\,s the majority of the molecules have been cooled into
the small area of phase space near the origin, but some
molecules remain distributed throughout the phase-space
acceptance. The distribution has separated into two components,
one cold and one hot. After 10\,s, 90\% of the molecules are in
the cold component, the remaining hot molecules form a halo in
phase space around the cold ones, and there is a region in
between where there are no molecules at all. The molecules that
are slow to cool are the ones that initially have large angular
momentum about the trap centre. In the absence of collisions,
these molecules cannot reach the centre, and if they cannot
reach the centre they are unlikely to have any collisions. The
molecules in the halo in Figure \ref{microwavePSDs}(d) have
particularly large angular momenta and so spend all their time
in the far wings of the atomic distribution; they are unlikely
to have collisions even after 10\,s.

\begin{figure}[t]
\includegraphics[width=8cm]{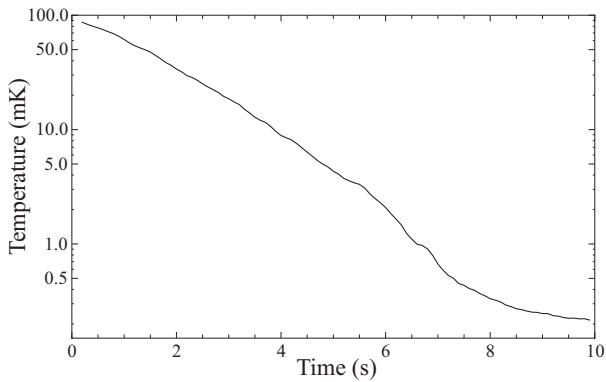}
\caption{
Temperature versus time for molecules thermalizing with cold
atoms in a microwave trap. The atom distribution has a
temperature of $T_a=140\,\mu$K and a width of $w_a=5$\,mm. To
obtain the molecule temperature, we take the kinetic energy
distribution, remove high energy outliers, take the mean and
divide by $\frac{3}{2}k$.} \label{TemperatureVsTime}
\end{figure}

The two-component speed distribution is even more evident when
the atom cloud width is reduced to 1\,mm. In this case, a cold
distribution develops rapidly at the centre of the trap but the
rest of the trap phase-space acceptance is filled with hot
molecules, apart from a thin empty region separating the hot
and cold distributions. When the atom cloud width is
$w_a=5$\,mm the atom-molecule overlap is sufficient that the
molecules form a single-component speed distribution. In this
case it is possible to give a sensible measure of the
temperature. The mean kinetic energy is not a good measure
because a few remaining outliers with high kinetic energy have
a disproportionate effect on the mean. Instead, we trim the
distribution by removing the 5\% that have the highest kinetic
energy, take the mean, and then divide by $\frac{3}{2}k$ to
obtain a temperature. Figure \ref{TemperatureVsTime} shows the
result. After 10\,s the molecules have cooled from an initial
temperature of 100\,mK to a final temperature of 200\,$\mu$K,
close to the temperature of the atom cloud. The mean number of
collisions per molecule required to reach this temperature is
30. As the molecules cool they move into the densest part of
the atomic distribution and, as shown in Figure \ref{fig:cc},
the collision cross section tends to increase. As a result, the
cooling rate tends to increase gradually between 100\,mK and
1\,mK, despite the fact that the collision rate is proportional
to the decreasing speed.

\begin{figure}[t]
\includegraphics[width=8cm]{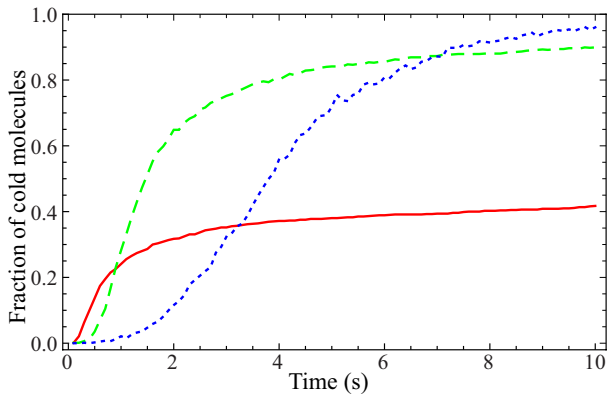}
\caption{(Color online)
Fraction of cold molecules in the simulated microwave trap as a function
of time, for three different atom cloud sizes: 1\,mm (red solid
line), 3\,mm (green dashed line) and 5\,mm (blue dotted line).
A molecule is classified as cold if its kinetic energy is less than
$\frac{3}{2} k T$, where $T=1$\,mK.} \label{coldMolsVsTime}
\end{figure}

Figure \ref{coldMolsVsTime} shows the number of cold molecules
as a function of time for three different atomic cloud sizes.
The number of molecules is normalized to the total number in
the trap, and a molecule is taken to be cold once its kinetic
energy is less than $\frac{3}{2} k T$, where $T=1$\,mK. When
$w_a= 5$\,mm, almost all the molecules in the trap cool to low
temperature, but the cooling is slow because the atom density
is low. After 1\,s, only 2\% of the molecules are cold, but
after 10\,s, 96\% of them are cold. As the atomic density is
increased by reducing the size of the cloud, the cooling rate
at early times increases. However, the number of cold molecules
obtained after a long period of cooling is lower with these
smaller atom clouds. For both the 1\,mm and 3\,mm atom clouds,
about 25\% of the molecules are cold after 1\,s. After 10\,s,
90\% of the molecules are cold when $w_a=3$\,mm, but only 42\%
when $w_a=1$\,mm. As discussed above, the molecules that fail
to cool are those that have large angular momentum about the
centre of the trap. These results show that the most suitable
choice of atom cloud size depends on the trap lifetime. If the
lifetime is long enough, it is best to use a large atom cloud
to maximize the number of cold molecules obtained. If the
lifetime is short, it is better to use a small atom cloud to
maximize the cooling rate and then to remove the molecules that
remain hot, for example by lowering the trap depth.

\section{Conclusions} \label{sec:conclusion}

Molecules are most easily trapped when they are prepared in
weak-field-seeking states, but then sympathetic cooling is
feasible only if the ratio of elastic to inelastic cross
sections is high. In the Li+LiH system with LiH in its
rotationally excited state ($j=1$), we find this ratio to be
approximately 5 at a collision energy of 100\,mK, gradually
falling as the collision energy decreases and reaching 1 at
about 30\,$\mu$K. This ratio is too small for sympathetic
cooling to be effective, since cooling from 100\,mK to
100\,$\mu$K results in 4 orders of magnitude of trap loss. To
avoid inelastic losses in this system, the molecules need to be
trapped in the ground state. This can be done using an ac
electric trap, but in this trap collisions can transfer stable
molecules onto unstable trajectories. This occurs because
stability of motion in an ac trap relies on a specific
correlation between position and speed, and this necessary
correlation tends to be upset by collisions. Our simulations
suggest that the resulting trap losses are too great for
sympathetic cooling to be feasible in a realistic ac trap.

Alternatively, ground-state molecules can be trapped in the
electric field maximum of a standing wave microwave field
formed inside a microwave cavity. The microwave trap appears to
be suitable for sympathetic cooling. We find that both the
cooling rate and the fraction of molecules that cool depend on
the degree of overlap between the atom and molecule
distributions. When the atoms are compressed to a small volume,
the cooling rate is high at the trap centre, but a large
fraction of the molecules do not cool because their angular
momentum prevents them from reaching the centre of the trap.
When the atom cloud is larger, more of the molecules cool but
the cooling is slower. The typical time required for a large
fraction of the molecules to reach ultracold temperatures is a
few seconds. For ground-state LiH, in a good vacuum, the trap lifetime will be limited by black-body heating of the rotational motion \cite{Hoekstra07}. This black-body-limited lifetime is 2.1\,s at room temperature, rising to 9.1\,s at 77\,K \cite{Buhmann08}. This suggests that liquid nitrogen cooling of the microwave cavity may be necessary in order to cool a large fraction of the molecules in the time available. Cooling of the cavity mirrors would also allow for a factor of 10 increase in the cavity Q-factor, and a corresponding decrease in the power required to obtain the same trap depth. Note that the black-body heating rate is considerably slower for many other molecules of interest \cite{Hoekstra07, Buhmann08}, and then trap lifetimes of several tens of seconds should be attainable under good vacuum conditions.

Our use of a hard-sphere scattering model may underestimate the degree of forward scattering, thereby overestimating the cooling rate since scattering in the forward direction does little to cool the molecules. We will investigate this in future work.  We have used an unchanging atomic distribution in our
simulations, but it is clear that the atoms could be used more
efficiently. It would be better to compress the atom cloud
gradually so that the size of the atom distribution matches
that of the molecules as they cool towards the trap centre.
This optimizes the collision rate by optimizing the atom
density and the overlap between the two clouds at all times.
Compressing the atom cloud will raise its temperature, but at
high densities the atoms can be evaporatively cooled which will
in turn sympathetically cool the molecules to even lower
temperatures. We have focussed on LiH molecules sympathetically
cooled with Li atoms, but we expect our general conclusions to
apply to a wide range of other systems.

\acknowledgments We are grateful to Isabel Llorente-Garcia and
Benoit Darqui\'e for helpful discussions regarding the
collision simulations. We acknowledge financial support from
the Polish Ministry of Science and Higher Education (grant
1165/ESF/2007/03) and from EPSRC under collaborative project
CoPoMol of the ESF EUROCORES Programme EuroQUAM.


\end{document}